\documentclass[aps,pra,reprint,superscriptaddress]{revtex4-1}

\usepackage{graphicx}
\usepackage{amsmath}
\usepackage{amsthm}
\usepackage{dsfont}
\usepackage{xcolor}

\begin{document}

\title{Quantum many-body scars with chiral topological order in 2D\\
and critical properties in 1D}

\author{N. S. Srivatsa}
\affiliation{Max-Planck-Institut f\"{u}r Physik komplexer Systeme, D-01187 Dresden, Germany}

\author{Julia Wildeboer}
\affiliation{Department of Physics, Arizona State University, Tempe, AZ 85287, USA}

\author{Alexander Seidel}
\affiliation{Department of Physics, Washington University, St. Louis, Missouri 63130, USA}

\author{Anne E. B. Nielsen}
\altaffiliation{On leave from Department of Physics and Astronomy, Aarhus University, DK-8000 Aarhus C, Denmark}
\affiliation{Max-Planck-Institut f\"{u}r Physik komplexer Systeme, D-01187 Dresden, Germany}

\begin{abstract}
We construct few-body, interacting, nonlocal Hamiltonians with a quantum scar state in an otherwise thermalizing many-body spectrum. In one dimension, the embedded state is a critical state, and in two dimensions, the embedded state is a chiral topologically ordered state. The models are defined on slightly disordered lattices, and the scar state appears independent of the precise realization of the disorder. A parameter allows the scar state to be placed at any position in the spectrum. We show that the level spacing distributions are Wigner-Dyson and that the entanglement entropies of the states in the middle of the spectrum are close to the Page value. Finally, we confirm the topological order in the scar state by showing that one can insert anyons into the state.
\end{abstract}

\maketitle

\section{Introduction}

Thermalization in quantum many-body systems is encoded in individual eigenstates of a generic Hamiltonian, and such a mechanism is hypothesised under the name eigenstate thermalization \cite{Deu,Sred}. Integrable systems, on the other hand, fail to thermalize owing to the existence of an extensive set of conserved quantities. Disordered systems also add to the list of counter examples to eigenstate thermalization. The disorder can  give rise to an emergent integrability in the systems that leads to a phenomenon called many-body localization where all eigenstates in the spectrum are non-ergodic at large enough disorder strengths \cite{Anderson,Basko,Nand}.

Recently, an experiment in a Rydberg atomic chain witnessed a weak ergodicity breaking where long time oscillations of local observables persisted when the system was initialised in a specific quantum many-body state \cite{Bernien} and such periodic revivals were termed ``quantum many-body scars" \cite{Turner,Turner2018,Lin} in  analogy to the quantum scars in chaotic single particle systems \cite{Heller}.

Currently, quantum many-body scars are defined as instances of one or more states in the spectrum of a non-integrable Hamiltonian that violate the strong sense of eigenstate thermalization and several interesting models have been studied and constructed that support these nontrivial states in the sea of thermal states \cite{Bull,Ho,Sanjay1,Sanjay2,Sala,Kormos2017,odea,pakrouski,ren}. The scar states essentially have low and sub-extensive entanglement entropy, similar to gapped or critical ground states. Systematic constructions have even been proposed to allow for a scar state to have nonchiral topological order \cite{Neupert,wildeboer} and survive in a disordered setting \cite{Onsager}.

Here, we construct a few-body Hamiltonian on 2D lattices, which has a scar state with chiral topological order in the middle of the spectrum. We also construct a few-body Hamiltonian on 1D lattices with a critical scar state. For both models, a parameter allows us to place the scar state at any desired position in the spectrum. The models are defined on slightly disordered lattices in 1D and 2D, and the scar state in 2D is a lattice Laughlin state. We provide evidence that the spectra are thermal by showing that the level spacing distributions are Wigner-Dyson and that the entanglement entropies of the excited states are close to the Page value. The scar state, on the contrary, is non-ergodic and has a much lower entanglement entropy. We demonstrate the topological properties of the scar state in 2D by introducing anyons into the state.

Our construction starts from sets of operators $A_i$ and $B_i$ that annihilate the scar state $|\Psi_{\textrm{scar}}\rangle$ of interest, i.e.\
\begin{equation}
A_i|\Psi_{\textrm{scar}}\rangle=
B_i|\Psi_{\textrm{scar}}\rangle=0.
\end{equation}
We use these to construct the Hamiltonian
\begin{equation}
H=\sum_{i} A_i^{\dagger} A_i - \gamma \sum_{i} B_i^{\dagger} B_i,
\end{equation}
which has $|\Psi_{\textrm{scar}}\rangle$ as an exact zero energy eigenstate. The Hamiltonian $H$ can have both positive and negative eigenvalues, and the real parameter $\gamma$ can be used to adjust the position of the scar state relative to the range of the many-body spectrum. For the special case $\gamma=0$, we observe that $|\Psi_{\textrm{scar}}\rangle$ is the ground state. We expect this approach to work quite generally to construct scar models, although the thermal properties of the spectrum need to be checked in each case.

The paper is structured as follows. In Sec.\ \ref{sec:scar}, we give the exact expression for the scar states in 1D and 2D. In Sec.\ \ref{sec:Ham}, we construct few-body Hamiltonians for these scar states by exploiting operators that annihilate the scar state. The Hamiltonians can be defined on arbitrary lattices in 1D and 2D, and we describe the slightly disordered lattices that we use for the numerical computations. In Sec.\ \ref{sec:HamProp}, we first show that the scar states can be placed at desired positions in the spectra by adjusting a parameter in the Hamiltonians. We then compute the energy level spacing distributions and entanglement entropies to show that the spectra are thermal. In Sec.\ \ref{sec:properties}, we discuss the properties of the scar states on the disordered lattices. The scar state in 1D is critical, and we show that the scar state in 2D is topologically ordered by constructing well screened anyons. Sec.\ \ref{sec:conclusion} concludes the paper.

\section{Quantum scar wavefunction}\label{sec:scar}

The exact wavefunction for the quantum scar that we wish to embed in a many-body spectrum is given by
\begin{equation}\label{state}
|\Psi_{\textrm{scar}}\rangle=
\sum_{n_1,n_2,\ldots,n_N}
\Psi(n_1, n_2, \ldots, n_N)
|n_1,n_2,\ldots,n_N\rangle,
\end{equation}
where
\begin{multline}
\Psi(n_1, n_2, \ldots, n_N)\propto\\
(-1)^{\sum_j(j-1)n_j} \; \delta_n \; \prod_{i<j}
(z_i - z_j)^{2n_in_j-n_i-n_j}.\nonumber
\end{multline}
Here, $n_i\in \{0,1\}$ is the number of hardcore bosons on the $i$th lattice site, and
\begin{eqnarray}
\delta_n &=&
\left\{\begin{array}{ll}
\displaystyle
1 & \mbox{for } \sum_i n_i= N/2
\\ [1ex]
\displaystyle
0 & \mbox{otherwise}
\end{array}\right.
\end{eqnarray}
fixes the number of particles in the state to $N/2$. The complex number $z_i$ denotes the position of the $i$th lattice site. In 1D, when the lattice sites sit evenly on a circle ($z_j=e^{i2\pi j/N}$), the state coincides with the ground state of the Haldane-Shastry model after a transformation to the spin basis \cite{HSmodelH,shast}. In 2D, on a square lattice, the state is the half filled bosonic lattice Laughlin state, which has chiral topological order \cite{Nielsen_2012}.

\section{Hamiltonians}\label{sec:Ham}

We now construct few-body Hamiltonians for which $|\Psi_{\textrm{scar}}\rangle$ is an exact eigenstate. The construction relies on the observation that the state \eqref{state} can be written as a correlation function in conformal field theory, and that one can use this property to derive families of operators that annihilate the state. The operators that we use below were derived in \cite{Tu}.

\subsection{Hamiltonian in 1D}

We first consider a 1D system, which is obtained by restricting all the positions $z_j$ to be on the unit circle, i.e.\ $z_j=e^{i\phi_j}$ for all $j$, where the $\phi_j$ are real numbers. We also require that all the $z_j$ are different from one another. It was shown in \cite{Tu} that the operators
\begin{equation}
\begin{split}
&\Lambda^{\textrm{1D}}_i=\sum_{j(\neq i)}w_{ij}[d_j-d_i(2n_j-1)],\\
&\Gamma^{\textrm{1D}}_i=\sum_{j(\neq i)}w_{ij}d_id_j, \qquad
w_{ij}=\frac{z_i+z_j}{z_i-z_j},
\end{split}
\end{equation}
annihilate $|\Psi_{\textrm{scar}}\rangle$. The operators $d_j$, $d^{\dagger}_j$, and $n_j=d^{\dagger}_j d_j$, acting on site $j$, are the annihilation, creation, and number operators for hardcore bosons, respectively.

We build the Hamiltonian for the scar model in 1D by constructing the Hermitian operator
\begin{equation}
H_{\textrm{1D}}=\sum_{i}{(\Lambda^{\textrm{1D}}_i)}^{\dagger}(\Lambda^{\textrm{1D}}_{i})
+(\alpha-2)\sum_{i}{(\Gamma^{\textrm{1D}}_{i})}^{\dagger}(\Gamma^{\textrm{1D}}_{i}),
\end{equation}
which has $|\Psi_{\textrm{scar}}\rangle$ as an exact energy eigenstate with energy zero. The real parameter $\alpha$ is chosen to position the scar state at the desired position in the spectrum of $H_{\textrm{1D}}$ as we shall see below. Expanding this expression, we get
\begin{multline}\label{ham1}
H_{\textrm{1D}}=\sum_{i\neq j}G^{A}_{ij} \, d_{i}^{\dagger}d_{j}
+\sum_{i\neq j}G_{ij}^{B}\, n_{i}n_{j}+\sum_{i\neq j \neq l}G_{ijl}^{C} \, d^{\dagger}_{i}d_{l}n_j\\
+\sum_{i}G^{D}_{i}\, n_i+G^{E},
\end{multline}
where the coefficients are given by
\begin{equation}
\begin{split}
&G^{A}_{ij}=-2w_{ij}^2,\\
&G^{B}_{ij}=(2-\alpha)w_{ij}^2+4\sum_{l(\neq j\neq i)}w_{ij}w_{il},\\
&G^{C}_{ijl}=-\alpha w_{ji}w_{jl},\\
&G^{D}_{i}=-2\sum_{j(\neq i)}w_{ij}^2-\sum_{j\neq l(\neq i)}w_{ij}w_{il},\\
&G^{E}=\frac{-N(N-2)(N-4)}{6}.
\end{split}
\end{equation}
Eq.\ \eqref{ham1} is thus a real, particle-number conserving 2-body Hamiltonian with some non-local terms.

As we aim at constructing a model with a thermal spectrum, we add a small amount of random disorder to the lattice site positions to avoid any additional symmetries. Specifically, we choose
\begin{equation}\label{zdisorder}
z_{j}=e^{2\pi i(j+\gamma_{j})/N}, \qquad
j\in\{1,2,\ldots,N\},
\end{equation}
where $\gamma_{j}$ is a random number chosen with constant probability density in the interval $[-\case{\delta}{2},\case{\delta}{2}]$ and $\delta$ is the disorder strength. In all the numerical computations below, we choose $\delta=0.5$ and consider one particular disorder realization.

\subsection{Hamiltonian in 2D}

We next allow the $z_j$ to be in the complex plane, and we assume that all the $z_j$ are different from one another. In this case, the operators
\begin{equation}
\begin{split}
&\Lambda^{\textrm{2D}}_i=\sum_{j(\neq i)}c_{ij}[d_j-d_i(2n_j-1)],\\
&\Gamma^{\textrm{2D}}_i=\sum_{j(\neq i)}c_{ij}d_id_j, \qquad c_{ij}=\frac{1}{z_i-z_j},
\end{split}
\end{equation}
annihilate $|\Psi_{\textrm{scar}}\rangle$ as shown in \cite{Tu}, and we use these to construct the Hamiltonian
\begin{equation}
H_{\textrm{2D}}=\sum_{i}{(\Lambda^{\textrm{2D}}_i)}^{\dagger}(\Lambda^{\textrm{2D}}_{i})
+\beta\sum_{i}{(\Gamma^{\textrm{2D}}_{i})}^{\dagger}(\Gamma^{\textrm{2D}}_{i})
\end{equation}
for the scar model in 2D. In this expression, $\beta$ is a real parameter, which we can adjust to position the scar state anywhere in the spectrum as we shall see below. After expanding out the terms, we get
\begin{multline}\label{ham2}
H_{\textrm{2D}} = \sum_{i \ne j} F^A_{ij} \, d^{\dagger}_i d_j
+ \sum_{i \ne j} F^B_{ij} \, n_i n_j
+ \sum_{i \ne j \ne k} F^C_{ijk} \, d^{\dagger}_i d_jn_k \\
+ \sum_{i \ne j \ne k} F^D_{ijk} \, n_i n_jn_k
+ \sum_{i} F^E_{i} \, n_i,
\end{multline}
where
\begin{align}
& F^A_{ij} = 2c^*_{ij}c_{ij} + \sum_{k (\ne i, \ne j)}(c^*_{ki}c_{kj} + c^*_{ji}c_{jk} + c^*_{ik}c_{ij}), \nonumber\\
& F^B_{ij} = \beta c^*_{ij}c_{ij} - 2\sum_{k (\ne i,\ne j)}(c^*_{ij}c_{ik} + c^*_{ik}c_{ij}), \nonumber \\
& F^C_{ijk} = -2(c^*_{ji}c_{jk} + c^*_{ik}c_{ij})+\beta c^*_{ki}c_{kj}, \nonumber \\
& F^D_{ijk} = 4 c^*_{ik}c_{ij}, \nonumber \\
& F^E_{i} = \sum_{j(\ne i)} (c^*_{ji}c_{ji} + c^*_{ij}c_{ij}) + \sum_{j,k (\ne i)}c^*_{ij}c_{ik}. \nonumber
\end{align}
The Hamiltonian conserves the number of particles, it is nonlocal, and it consists of terms involving up to three particles.

We disorder the positions of the lattice sites slightly to avoid additional symmetries in the model. Specifically, if $z^{c}_{i}$ is the position of the $i$th site in the perfect square lattice, we choose $z_{i}=z^{c}_{i}+\delta(\epsilon_a+i\epsilon_b)$ where $\epsilon_a$ and $\epsilon_b$ are independent random numbers chosen with constant probability density in the interval $[0,1]$ and $\delta$ is the disorder strength. In all the numerical computations below, we choose $\delta=0.1$ and consider one particular disorder realization.

\section{Properties of the Hamiltonians}\label{sec:HamProp}

Since the Hamiltonians conserve the number of particles, and the scar state has $N/2$ particles, where $N$ is the number of lattice sites, we restrict our focus to the sector with lattice filling one half in the rest of the article.

\subsection{Positioning the scar state}

\begin{figure}
\includegraphics[width=\columnwidth]{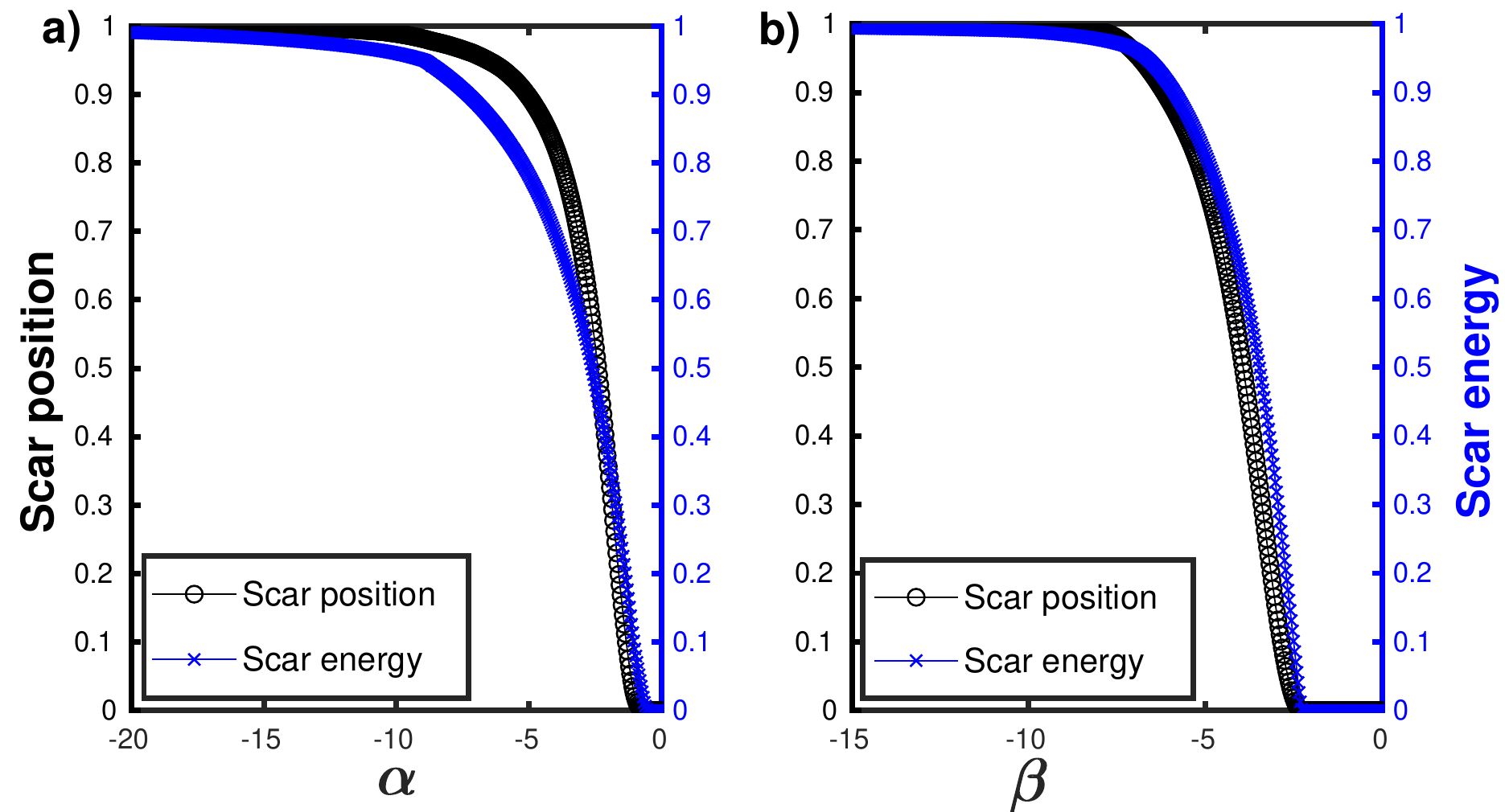}
\caption{(a) Normalized position \eqref{normpos} and normalized energy \eqref{normenergy} of the scar state in the many-body spectrum of the 1D Hamiltonian \eqref{ham1} plotted as a function of the tuning parameter $\alpha$ for a slightly disordered lattice on a unit circle with $N=16$ sites. (b) Same as (a), but computed for the 2D Hamiltonian \eqref{ham2} for a slightly disordered square lattice of size $4\times4$. It is seen that, both in 1D and 2D, the scar state can be positioned at any desired place in the many-body spectrum by a suitable choice of $\alpha$ and $\beta$, respectively.}\label{fig:position}
\end{figure}

We first show that the parameters $\alpha$ and $\beta$ serve as tuning parameters to place the scar state at the desired position of the many-body spectrum. If the scar state is the $n$th eigenstate, when ordering the eigenstates in ascending order according to energy, we define the normalized position of the scar state as
\begin{equation}\label{normpos}
\textrm{Normalized position}=\frac{n}{D},
\end{equation}
where $D={N \choose N/2}$ is the dimension of the Hilbert space in the half filled sector. The energy of the scar state is zero by construction, and the normalized energy of the scar state is hence
\begin{equation}\label{normenergy}
\textrm{Normalized energy}=\frac{-E_\textrm{min}}
{E_\textrm{max}-E_\textrm{min}},
\end{equation}
where $E_{\textrm{min}}$ ($E_{\textrm{max}}$) is the lowest (highest) energy in the spectrum of the Hamiltonian.

In Fig.\ \ref{fig:position}, we show these two quantities for the 1D and 2D Hamiltonians in Eqs.\ \eqref{ham1} and \eqref{ham2}, respectively. From the plots it is seen that the scar state can be placed at any desired position in the many-body spectrum by choosing $\alpha$ and $\beta$ appropriately. The scar state is the ground state for $\alpha=0$ and $\beta=0$, respectively.

\subsection{Level spacing distribution}

\begin{figure}
\includegraphics[width=\columnwidth]{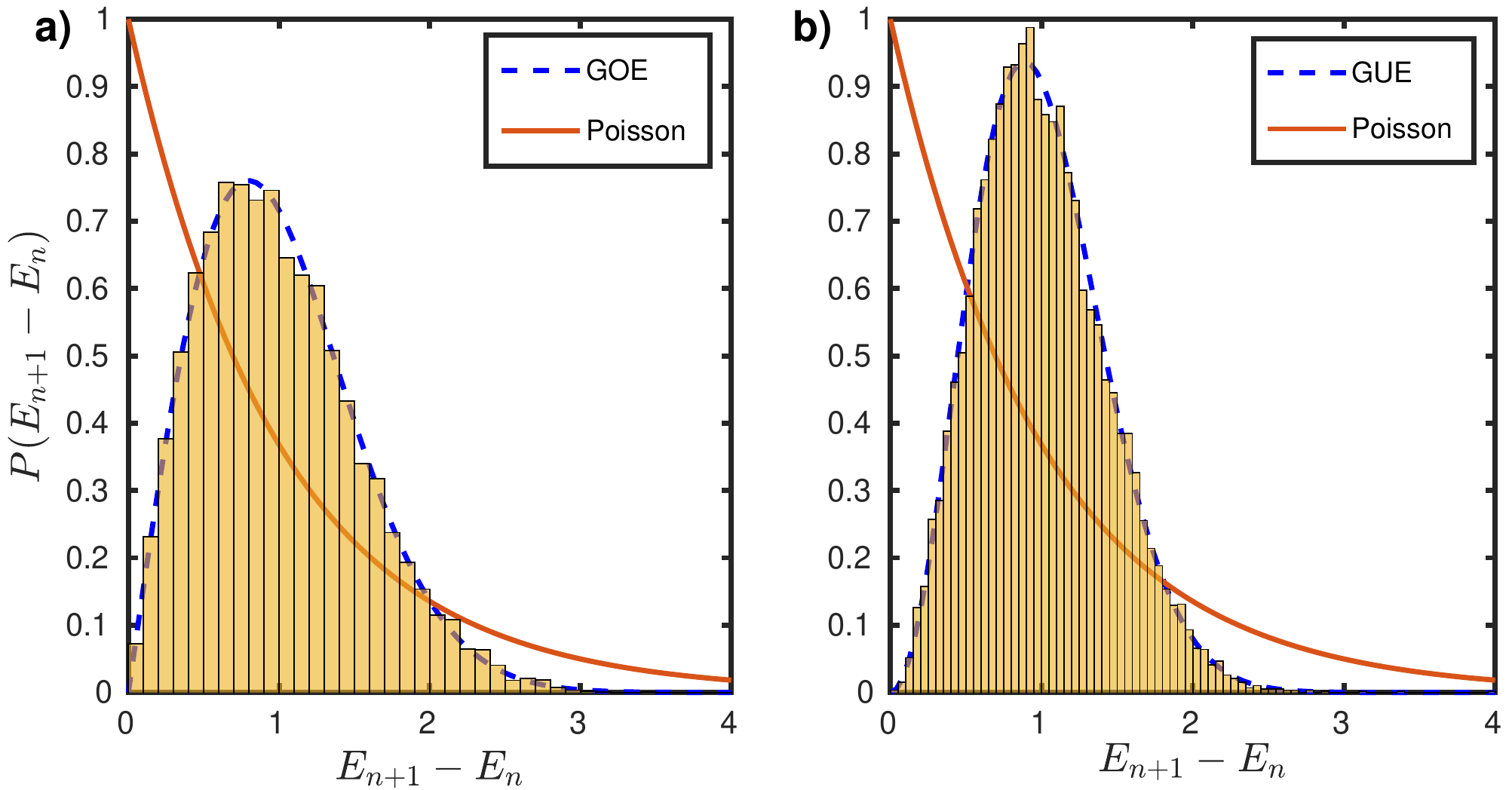}
\caption{(a) The level spacing distribution for the Hamiltonian in 1D with $\alpha=-4.5$ matches the GOE statistics for a system that is time-reversal invariant. (b) The level spacing distribution for the Hamiltonian in 2D with $\beta=-4$ matches the GUE statistics for a system that is not time-reversal invariant. This suggests that most of the states in the spectrum are ergodic.}\label{fig:levelspacing}
\end{figure}

We next show numerically that the level spacing distributions follow the Wigner surmise for ergodic Hamiltonians \cite{Wigner}. In particular, time reversal invariant Hamiltonians with real matrix entries are known to follow Gaussian orthogonal ensemble (GOE) statistics while those with complex entries breaking time reversal symmetry are known to follow Gaussian unitary ensemble (GUE) statistics \cite{Luca}. In Fig.\ \ref{fig:levelspacing}, we show the level spacing distributions for the Hamiltonians in 1D and 2D using the standard technique of unfolding the energy spectrum \cite{Takai}. The energy level spacing distribution follows GOE statistics for the 1D case while it follows GUE statistics for the 2D case. This suggests that most of the states in the many-body spectrum of the Hamiltonians are ergodic.

\subsection{Entanglement entropy}

Scar states are non-ergodic states that reside in the bulk of the spectrum where the density of states is high. The entanglement entropy of scar states is low compared to the entanglement entropy of the states in the middle of the spectrum that obey the eigenstate thermalization hypothesis. When the subsystem is half of the system, the entanglement entropy of the latter states is close to the Page value $[N\ln(2)-1]/2$ \cite{page}.

\begin{figure}
\includegraphics[width=\columnwidth]{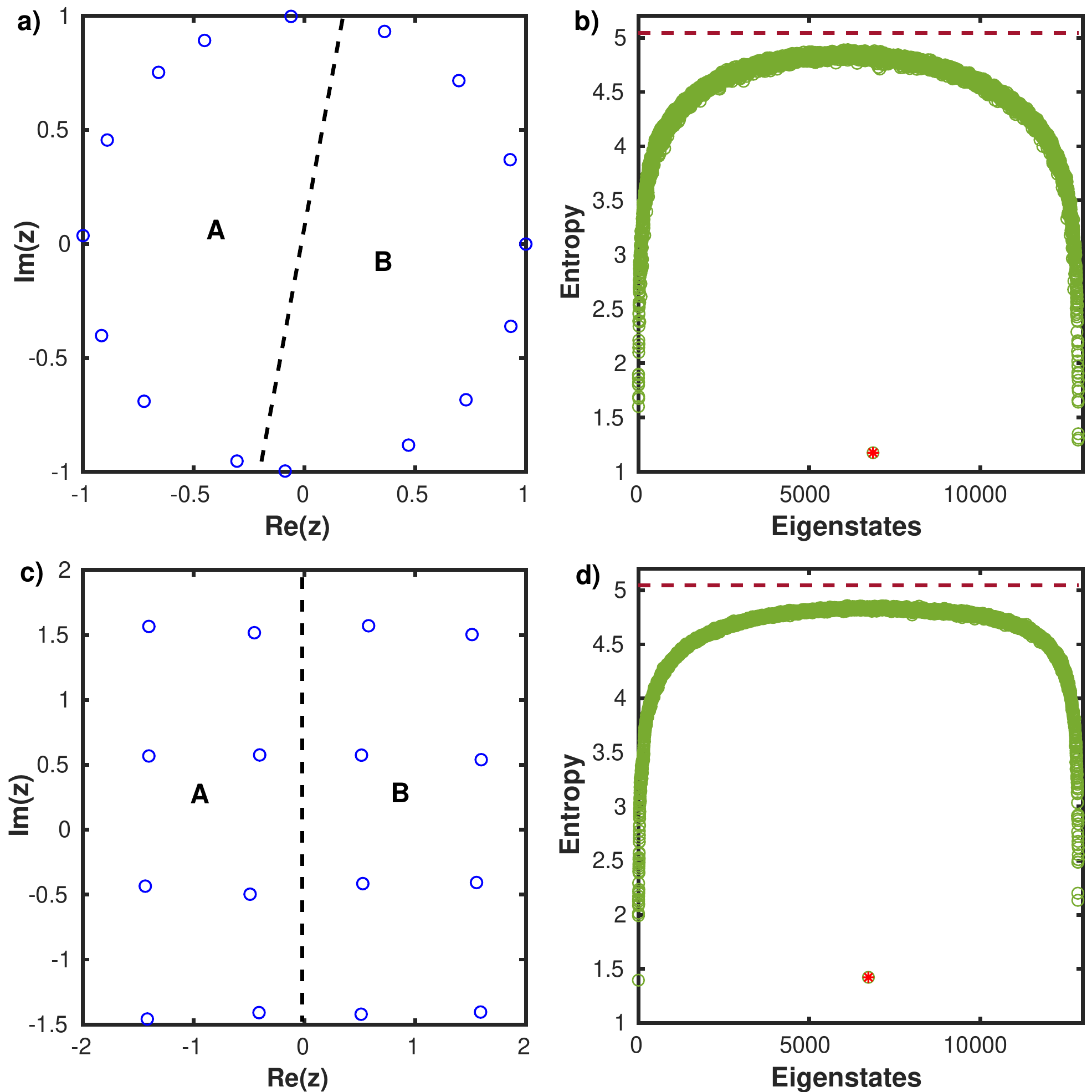}
\caption{(a,c) The slightly disordered lattices in 1D and 2D with $N=16$ spins. The red dashed lines show the partitioning into subsystems $A$ and $B$ used for computing the entanglement entropies. (b,d) Entanglement entropy for all the eigenstates of the Hamiltonian in 1D [Eq.\ \eqref{ham1}] with $\alpha=-4.5$ and in 2D [Eq.\ \eqref{ham2}] with $\beta=-4$, respectively. The states in the middle of the spectrum have entropies close to the Page value $[N\ln(2)-1]/2=5.05$ (dashed lines), except the scar state (marked with a red star), which has a much lower entropy.} \label{fig:entropy}
\end{figure}

We now show that the models considered here display such a behavior. Specifically, we consider the von Neumann entropy $S = -\mathrm{Tr}[\rho_A\ln(\rho_A)]$ of a subsystem $A$, where $\rho_A=\mathrm{Tr}_{B}(|\psi \rangle \langle \psi|)$ is the reduced density matrix after tracing out part $B$ of the system and $|\psi\rangle$ is the energy eigenstate of interest. For the 1D case, we choose region $A$ to be half of the chain, and for the 2D case, we choose region $A$ to be the left half of the lattice. The choice of subsystems and the entanglement entropy for all the states for a disordered case are shown in Fig.\ \ref{fig:entropy}. It is seen that the states in the middle of the spectrum have large entanglement entropy close to the Page value while the scar state has low entanglement entropy in both 1D and 2D. 

\begin{figure}
\includegraphics[width=\columnwidth]{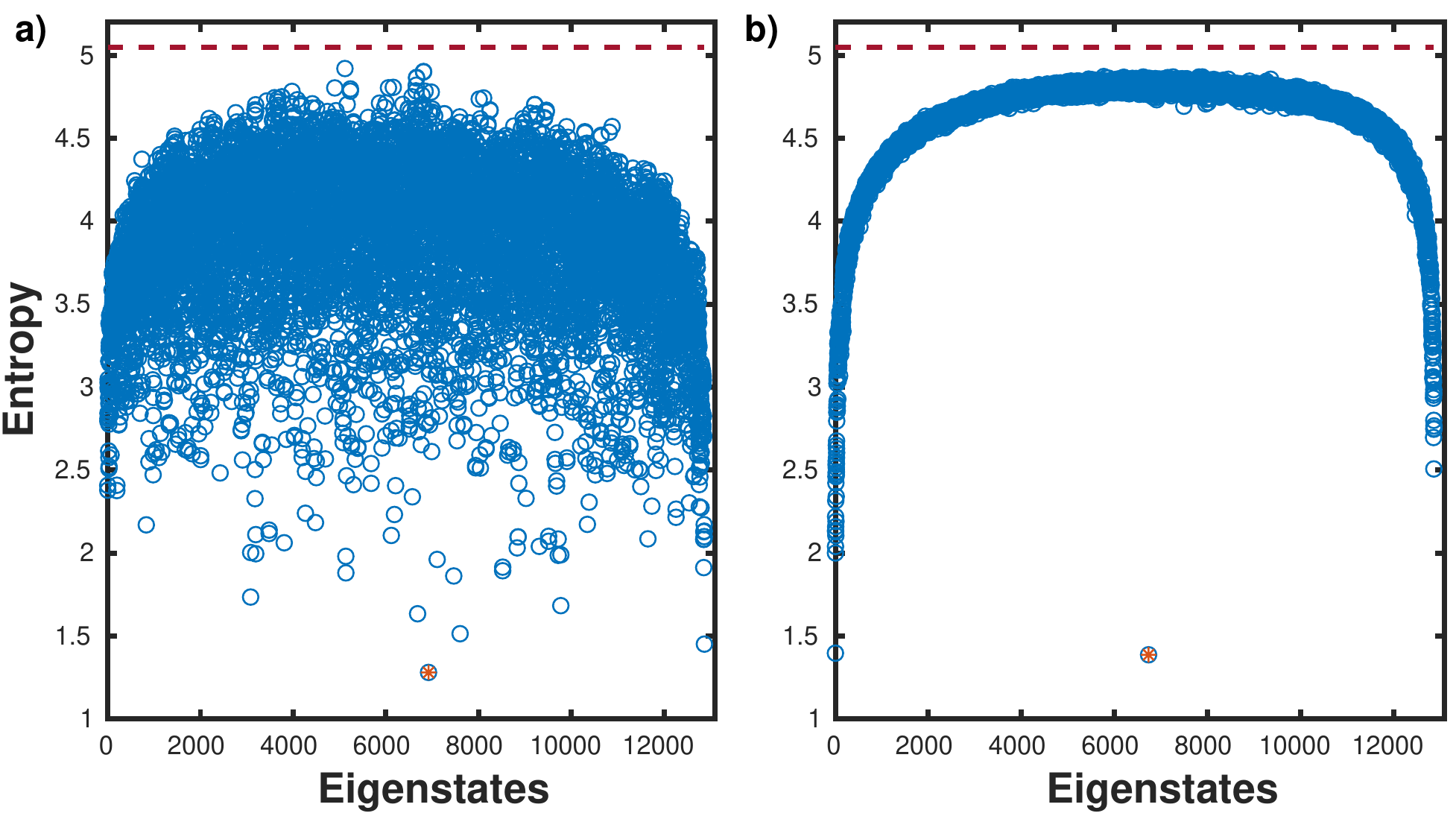}
\caption{(a) Entanglement entropy for all the eigenstates of the Hamiltonian in 1D [Eq.\ \eqref{ham1}] with $\alpha=-4.5$ and $\delta=0$. (b) Entanglement entropy for all the eigenstates of the Hamiltonian in 2D [Eq.\ \eqref{ham2}] with $\beta=-4$ and $\delta=0$. Dashed lines indicate the Page value $[N\ln(2)-1]/2=5.05$ and the scar state is marked with a red star.} \label{fig:entropyclean}
\end{figure}

We also compute entropies for models without disorder as shown in Fig.\ \ref{fig:entropyclean}. For the 1D case, the entropies in the middle of the spectrum are no longer distributed around the Page value. However, for the 2D clean model, the entropies close to the middle of the spectrum are close to the Page value while the scar state has low entanglement.

\section{Properties of the scar states}\label{sec:properties}

As mentioned above, the scar state coincides with the ground state of the Haldane-Shastry model, when defined on a uniform 1D lattice, and on a 2D square lattice, the scar state is a lattice Laughlin state. These properties remain the same after disordering the lattices slightly as we will now discuss.

\subsection{Critical state in 1D}

The ground state of the Haldane-Shastry model has been studied extensively and is known to be a critical state with Luttinger liquid properties \cite{HSmodelH,shast}. The entanglement entropy of a critical state scales logarithmically with the subsystem size $L$. More precisely \cite{Calabrese}
\begin{equation}
S\approx\frac{c}{3}\ln\left[\frac{N}{\pi}\sin\left(\frac{\pi L}{N}\right)\right]+\textrm{constant},
\end{equation}
where $c$ is the central charge of the underlying conformal field theory, and $c=1$ for the Haldane-Shastry state. Numerous studies have confirmed that the state is robust to weak disorder and does not change its universality class {\cite{cirac,stephan2016full}. The entropy of the scar state hence scales logarithmically also on the slightly disordered lattice.

\begin{figure}
\includegraphics[width=\columnwidth]{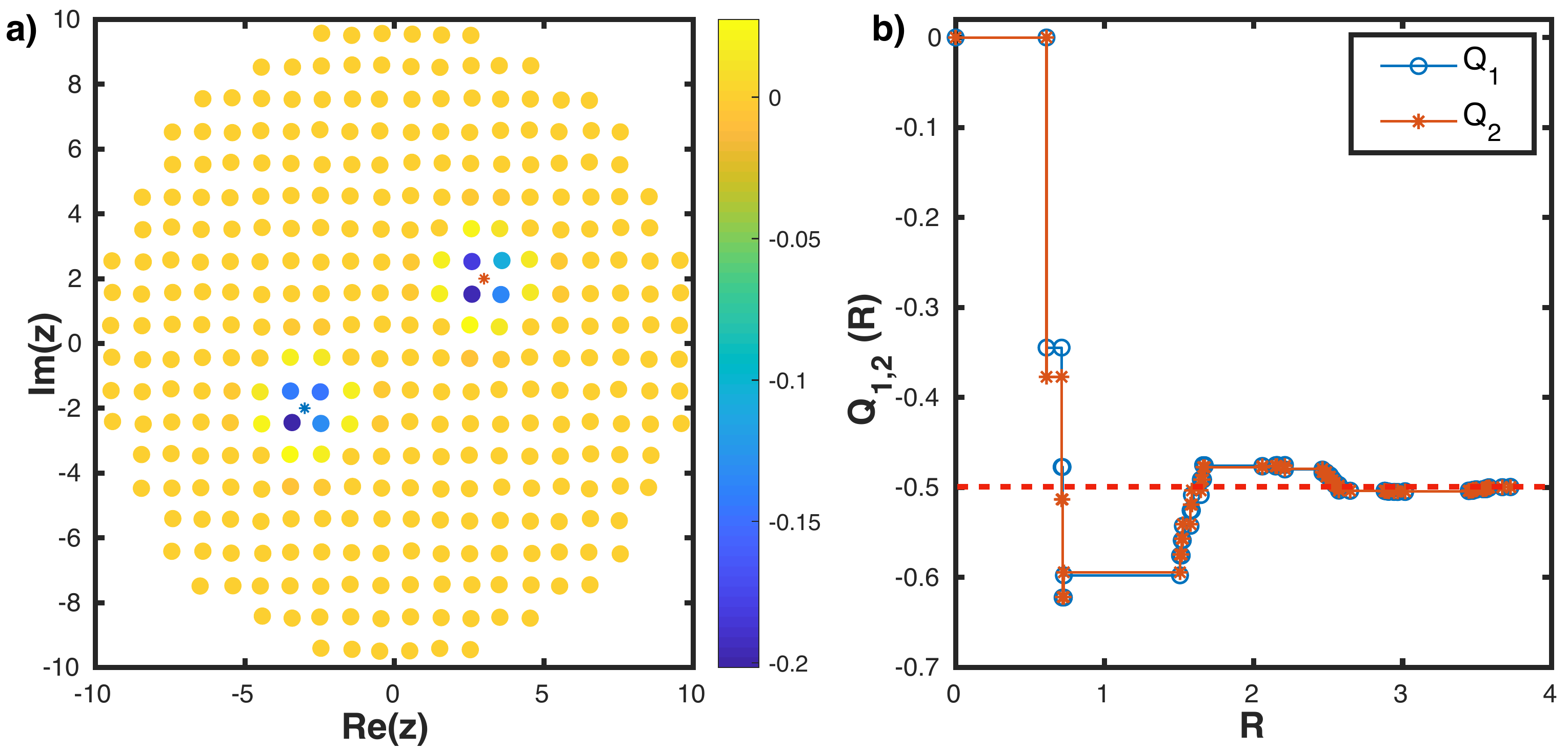}
\caption{(a) Density difference profile $\rho(z_i)$ [see \eqref{density}] on a slightly disordered square lattice with the two quasihole coordinates, $w_1$ and $w_2$, marked by the blue and the red star, respectively. The observation that $\rho(z_i)$ is only nonzero in a finite region around each of the quasihole coordinates shows that the anyons are well screened. (b) The excess densities, $Q_1$ and $Q_2$ [see \eqref{excesscharge}], as a function of the radius $R$ converge to the correct value $-0.5$ for a quasihole, which shows that the state is topological (see text). The number of lattice sites is $N = 316$ and the edge of the lattice is chosen to be roughly circular in order to resemble a quantum Hall droplet. The error bars due to Metropolis sampling are of order $10^{-4}$.}\label{fig:anyons}
\end{figure}

\subsection{Lattice Laughlin state in 2D}

The lattice Laughlin state \eqref{state} is known to have chiral topological order and to support anyonic excitations when defined on different regular lattices \cite{Nielsen_2012,Tu,anyon,ivan}. Here, we confirm the topological properties of the scar state on a slightly disordered lattice by introducing anyons. The anyons are introduced by modifying the state as follows \cite{anyon}.
\begin{equation}\label{PsiAnyon}
|\Psi_{\textrm{A}}\rangle=
 \sum_{n_1,\ldots,n_N}
 \Psi_{\textrm{A}}(w_1,w_2,n_1,\ldots,n_N)
 |n_1,\ldots,n_N\rangle
\end{equation}
where
\begin{multline}
\Psi_{\textrm{A}}(w_1,w_2,n_1,\ldots,n_N)=\\
C(w_1,w_2)^{-1}\;\delta_n\;(-1)^{\sum_j(j-1)n_j}
\prod_{i,j}(w_i-z_j)^{n_j}\\
\times\prod_{i<j}(z_i-z_j)^{2n_in_j}\prod_{i\neq j}(z_i-z_j)^{-n_i}.\nonumber
\end{multline}
Here, $w_1$ and $w_2$ are the centers of the two anyons, $C(w_1,w_2)$ is a real normalization constant, and
\begin{eqnarray}
\delta_n &=&
\left\{\begin{array}{ll}
\displaystyle
1 & \mbox{for } \sum_i n_i= (N-2)/2
\\ [1ex]
\displaystyle
0 & \mbox{otherwise}
\end{array}\right.
\end{eqnarray}
fixes the number of particles to $(N-2)/2$.

For a topological liquid, one expects a constant density of particles in the bulk. In the presence of quasiholes, however, the density is reduced in a local region around each of the quasihole coordinates $w_k$. We define the excess density of the $k$th quasihole as
\begin{equation}\label{excesscharge}
Q_k=\sum_{i=1}^{N}\rho(z_i)\,\Theta(R-|z_i-w_k|), \quad k\in\{1,2\},
\end{equation}
where
\begin{equation}\label{density}
\rho(z_i)=\langle\Psi_A| n_i |\Psi_A\rangle
-\langle \Psi| n_i |\Psi \rangle
\end{equation}
is the density difference between the situations when there are quasiholes in the system and when there are not. The Heaviside step function $\Theta(\ldots)$ restricts the sum to a circular region of radius $R$ and center $w_k$. When $R$ is large enough to enclose the quasihole and small enough to not enclose other quasiholes or part of the edge of the lattice, the excess density is equal to minus the charge of the quasihole. The charge of the quasihole is $1/2$ for the considered Laughlin state.

We use Metropolis-Hastings algorithm to compute $\rho(z_i)$ and $Q_k$ in Fig.\ \ref{fig:anyons}. The density difference profile shows that the quasiholes are well screened, and we find that the excess densities $Q_1$ and $Q_2$ converge to $-0.5$ as we increase the radius $R$. For the lattice Laughlin state, it can be shown analytically that the quasiholes have the correct braiding properties, if it can be assumed that the anyons are screened and have the correct charge \cite{Nielsen_2018}. Since the scar state hence admits the insertion of local defects with anyonic properties, it follows that it is a topologically ordered state even in the presence of the amount of disorder considered here.

\section{Conclusion}\label{sec:conclusion}

We have demonstrated the construction of 2D models featuring quantum many-body scar states that have chiral topological order. Specifically, we have used sets of operators that annihilate the scar state of interest to construct nonlocal, few-body Hamiltonians, which have the scar state as an exact eigenstate. Using a similar construction for a 1D system, we have also obtained a scar model in which the scar state is critical. For both models, a parameter allows us to place the scar state at any desired position in the spectrum. We have found that the level spacing distributions of the spectra are Gaussian and that the excited states in the middle of the spectra have entanglement entropies close to the Page value, which is what is expected for thermal spectra. The scar state, however, is special and has a much lower entanglement entropy. Finally, we have confirmed the topological nature of the scar state in 2D by showing that one can insert anyons with the correct charge and braiding properties into the state.

We expect that the construction method used can be applied to build several other scar models. One can, for instance, also find families of operators that annihilate Halperin states \cite{tu2014quantum} and Moore-Read states \cite{Glasser_2015}. Operators can also be found that annihilate, e.g., lattice Laughlin states with anyons \cite{anyon}. This would allow for the construction of scar models where the scar state contains anyons. In all cases, one should check, e.g.\ numerically, whether the spectra of the constructed models are, indeed, thermal. For the models mentioned, the lattice positions serve as free parameters, so even if the spectrum is not thermal for a particular choice of lattice, it may be thermal for another choice.

\begin{acknowledgments}
AS gratefully acknowledges support by the National Science Foundation under NSF Grant No. DMR-2029401.
\end{acknowledgments}


%

\end{document}